\documentclass{article}
\usepackage{ltwol2e}
\def \ts{\thinspace}

\def \GeV{{\rm \enspace GeV}}
\def \beq{\begin{equation}}
\def \eeq{\end{equation}}
\def \beqa{\begin{eqnarray}}
\def \eeqa{\end{eqnarray}}
 
\begin{document}

%%%%%%%%%%%%%%%%%%%%%%%%%%%%%%%%%%%%%%%%%%%%%%%%%%%%%%%%%%%%%%%%
%%
%%    TITLE AND ABSTRACT
%%
%%%%%%%%%%%%%%%%%%%%%%%%%%%%%%%%%%%%%%%%%%%%%%%%%%%%%%%%%%%%%%%%

\title{Theoretical Expectations in Radiative Top Decays}

\author{Gregory Mahlon$^{*}$}

\address{Department of Physics, McGill University, 
3600 University St., Montr\'eal, QC  H3A 2T8 Canada\\
Electronic address:  mahlon@physics.mcgill.ca}

\date{October 22, 1998}

\twocolumn[\maketitle\abstracts{
We summarize the Standard Model predictions for the three-body
decays of the top quark
$t \rightarrow WbX$, where $X=Z,H,g$ or $\gamma$.
Because of strong phase space suppression,
we find that the branching ratios for the $Z$ and $H$ final
states are of order a few times $10^{-7}$, rendering them
invisible at Tevatron Run II.  On the other hand,
the decays to $g$ and $\gamma$
are suppressed only by the expected factor of $\alpha_s$ or
$\alpha_{em}$. \hfill McGill/98-31}]

%%%%%%%%%%%%%%%%%%%%%%%%%%%%%%%%%%%%%%%%%%%%%%%%%%%%%%%%%%%%%%%%
%%
%%    INTRODUCTION -- INTRODUCTION
%%
%%%%%%%%%%%%%%%%%%%%%%%%%%%%%%%%%%%%%%%%%%%%%%%%%%%%%%%%%%%%%%%%

According to the Standard Model,
the decay $t\rightarrow Wb$
is by far the dominant decay mode of the top quark.  
Nevertheless, it is worthwhile to 
search for the other predicted decay modes of the top quark in
order to more completely test the Standard Model.
In this talk, we will examine the
3-body radiative decays $t\rightarrow W b X$, where $X$ can be a 
$Z$ boson, Higgs boson, gluon or photon. 

%%%%%%%%%%%%%%%%%%%%%%%%%%%%%%%%%%%%%%%%%%%%%%%%%%%%%%%%%%%%%%%%
%%
%%    t --> W b Z
%%
%%%%%%%%%%%%%%%%%%%%%%%%%%%%%%%%%%%%%%%%%%%%%%%%%%%%%%%%%%%%%%%%

An amusing coincidence involves
the masses of the top quark, the bottom quark, and the two
heavy gauge bosons.  We observe that
\beq
M_t \sim M_W + M_b + M_Z.
\eeq
Using the masses tabulated by the Particle Data Group,\cite{PDG}
this relation reads
\beq
173.8 \pm 3.5 \pm 3.9 \GeV \sim 176.1\pm 0.5\GeV,
\eeq
where the uncertainty on the right hand side is entirely due
to the ambiguity in the $b$-quark mass.
Thus, the on-shell decay of a top quark to a $WbZ$ final
state is on the verge of being allowed.  As a consequence,
effects of the finite width of the top quark become
crucial in calculating this decay rate.\cite{FiniteWidth}

In the stable-particle limit, there are three Feynman diagrams
which contribute to the decay $t\rightarrow W b Z$, since
the $Z$ may be radiated from any of the $t$, $W$, or $b$.
To account for the finite widths of the $W$ and $Z$ bosons,
we compute diagrams which include the decay products of the 
$W$ and $Z$, employing the Breit-Wigner form of the propagators
in the unitary gauge.  For example, we take the $W$ propagator
to be
\beq
{
{-i}
\over
{W^2-\hat m_W^2}
}
\biggl(g^{\mu\nu} -  
{
{{W^\mu}W^{\nu}}
\over
{\hat m_W^2}
}\biggr)
\eeq
where $\hat m_W \equiv m_W + {1\over2}i\Gamma_W$.
This form of the propagator respects the necessary gauge
invariance of the $t \rightarrow W b \gamma$ amplitude.

To simplify the computation, we consider only the decays
$W \rightarrow \mu^{+} \nu_\mu$ and $Z \rightarrow e^{+} e^{-}$,
and extract the rate by dividing by the appropriate branching
fractions:
\beq
\Gamma(t\rightarrow W b Z) =
{ {\Gamma(t\rightarrow \mu^{+}\nu_\mu b e^{+} e^{-})}
\over
{ {\cal B}(W\rightarrow \mu^{+}\nu_\mu)
  {\cal B}(Z\rightarrow e^{+}e^{-}) }
}.
\eeq
In all, a total of nine Feynman diagrams must be
considered.

                %%%%%%%%%%%%%%%%%%%%%%%%%%%%
                %%                        %%
                %%        FIGURE 1        %%
                %%                        %%
                %%%%%%%%%%%%%%%%%%%%%%%%%%%%
\begin{figure}
\vskip10.0cm
\includegraphics{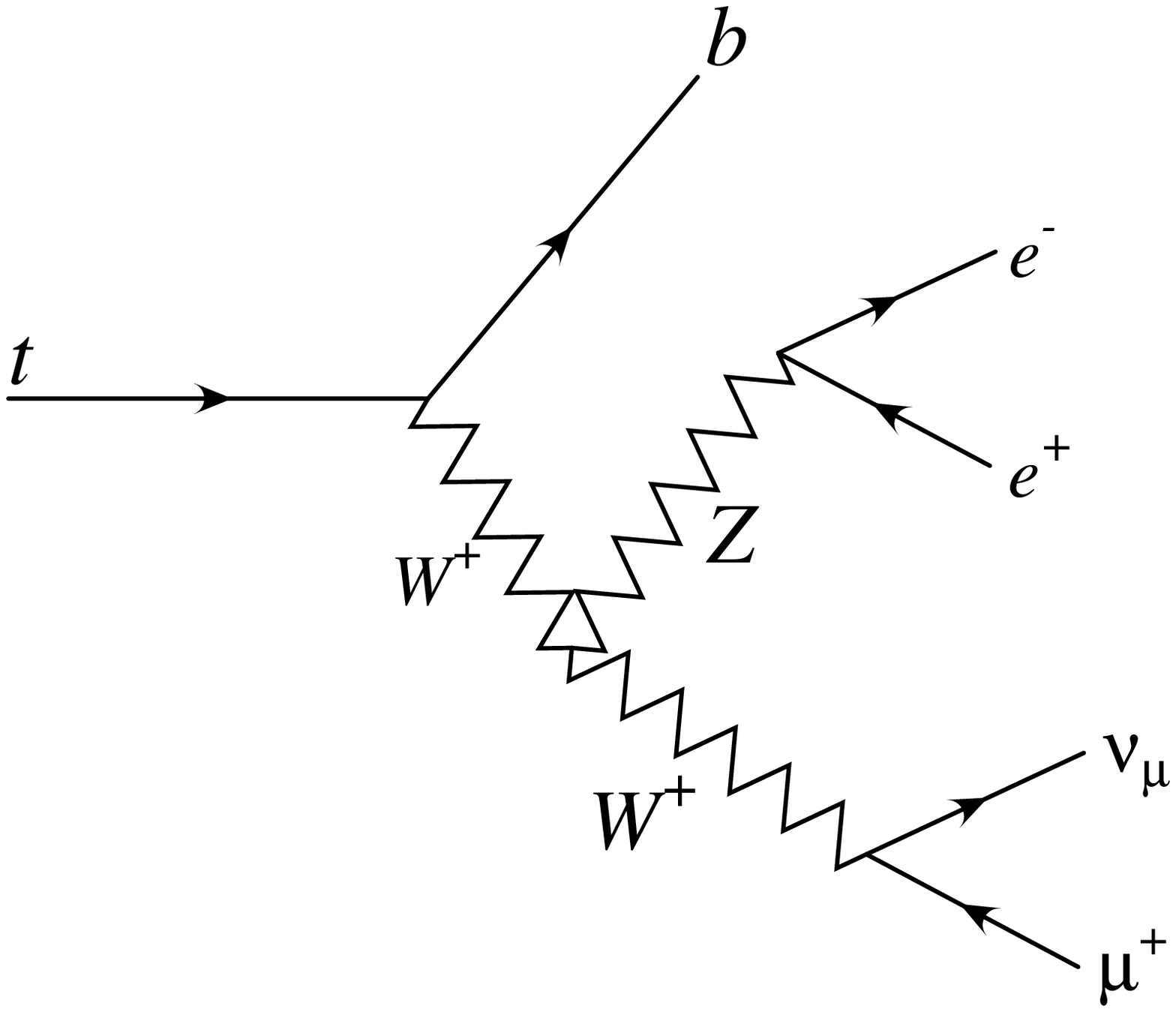}
\includegraphics{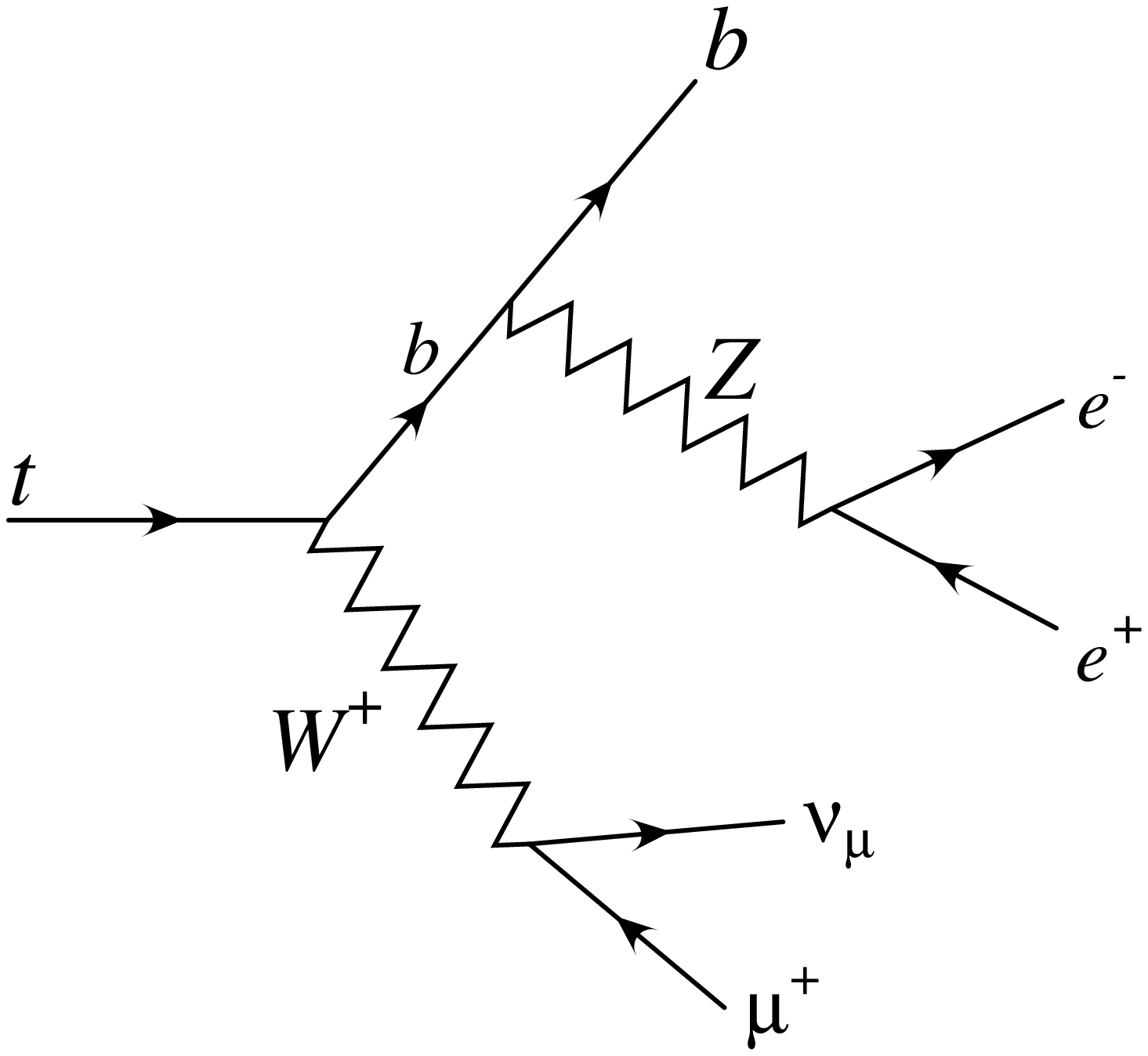}
\includegraphics{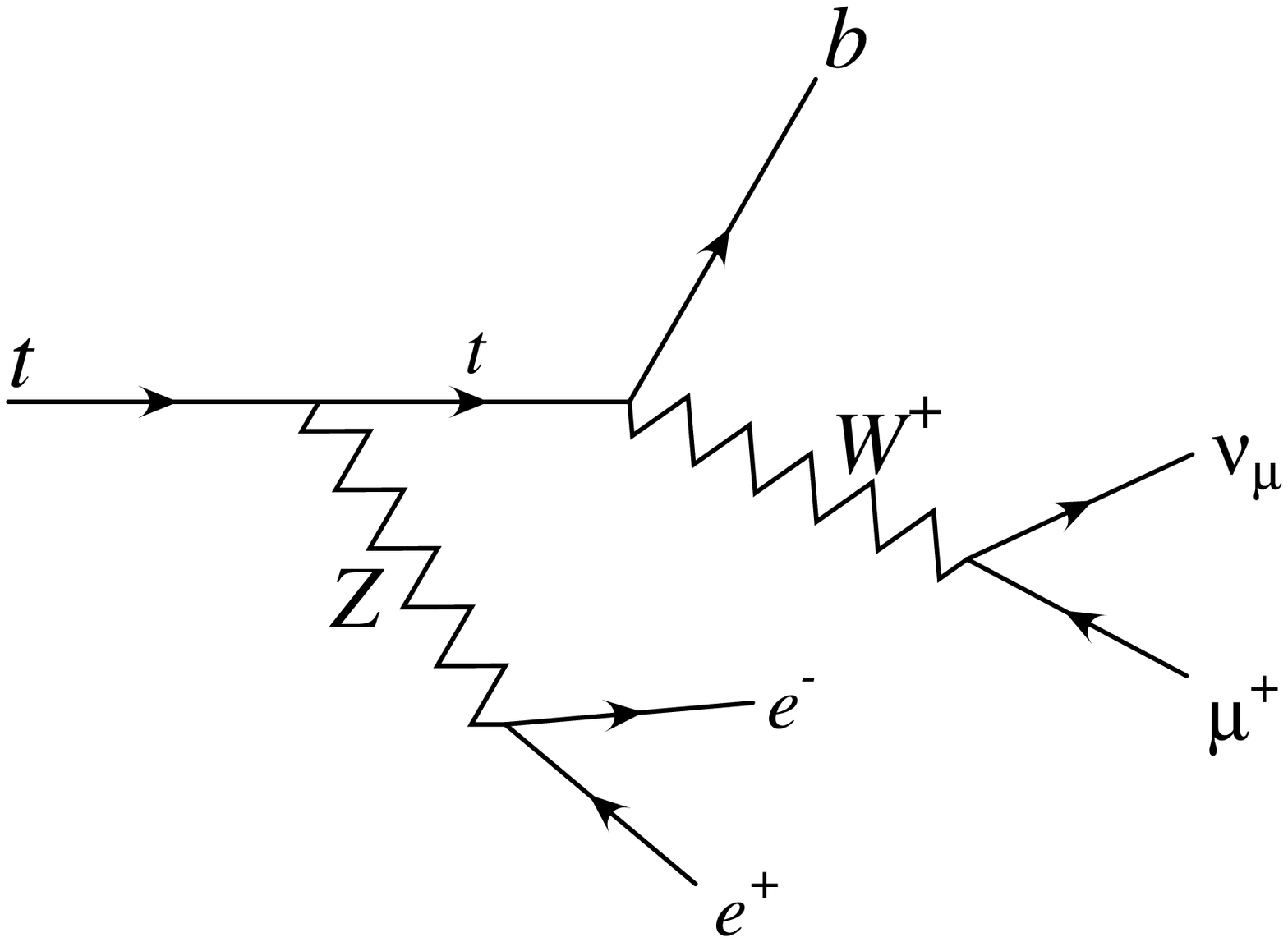}
\vspace*{-5cm}
\caption[]{Resonant contributions to the process
$t \rightarrow WbZ \rightarrow \mu^{+}\nu_\mu b e^{+}e^{-}$.
For large enough top mass, both the $W$ and $Z$ bosons are on shell.}
\label{resonant}
\end{figure}

\begin{figure}
\vskip10.0cm
\includegraphics{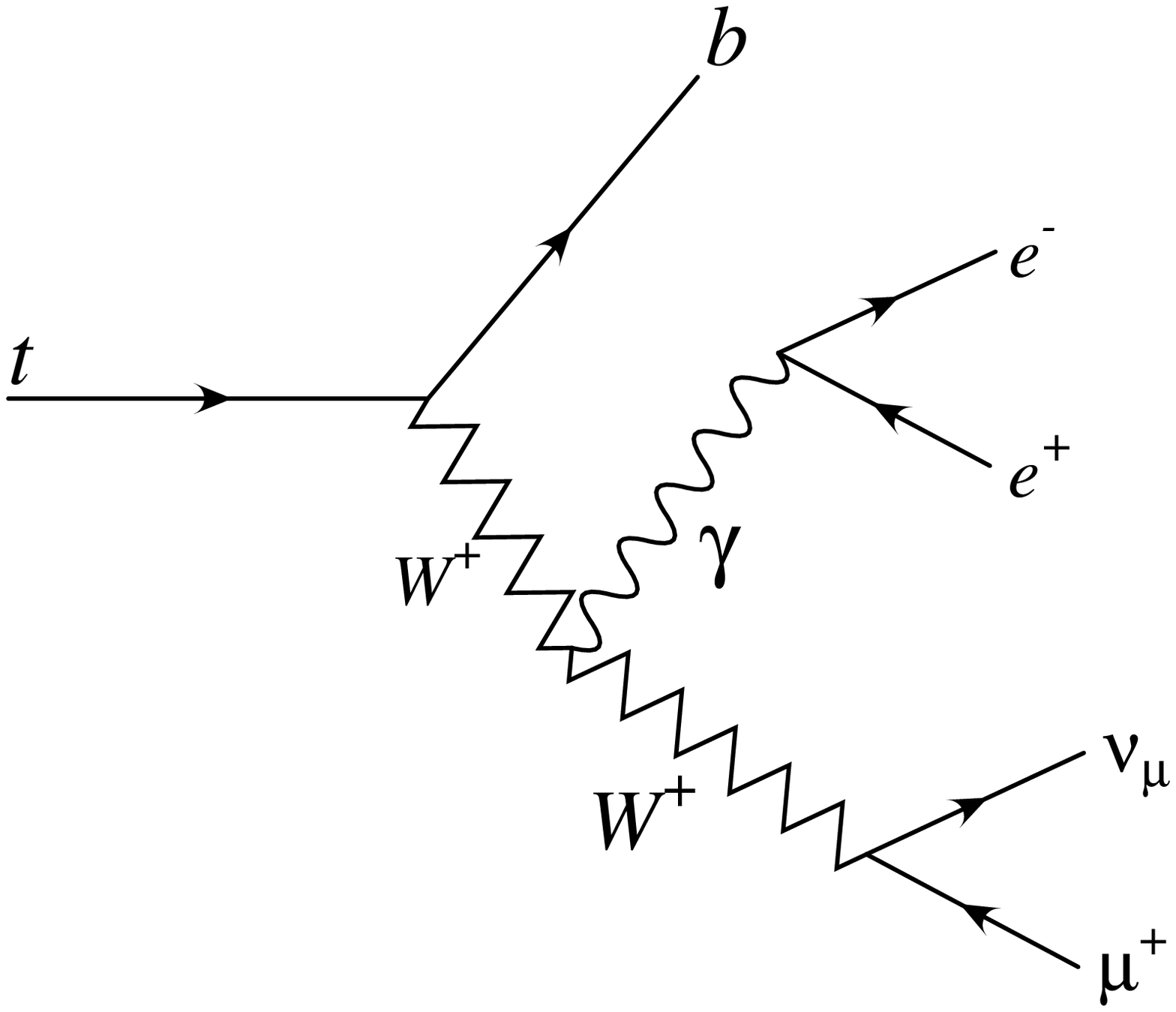}
\includegraphics{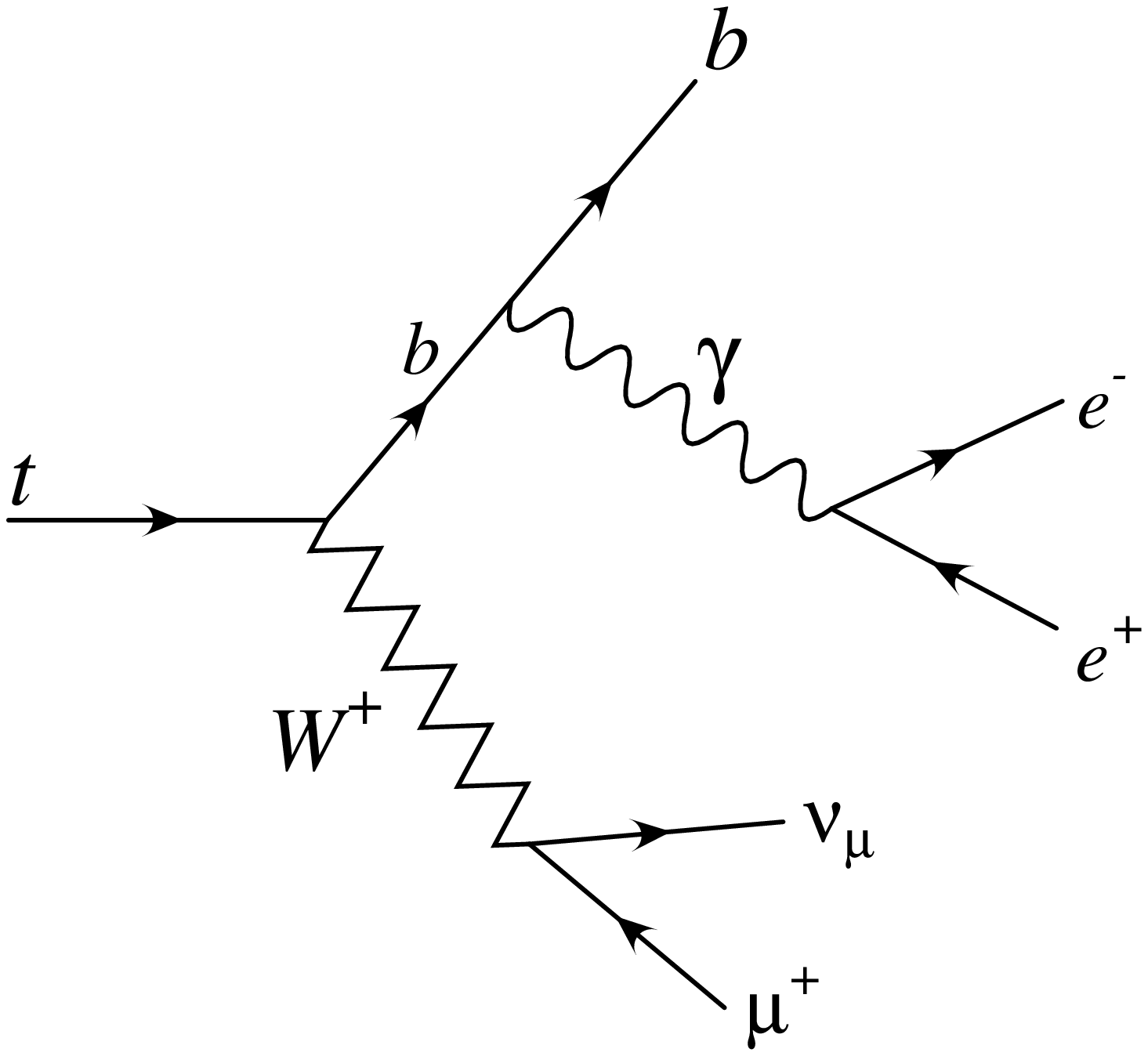}
\includegraphics{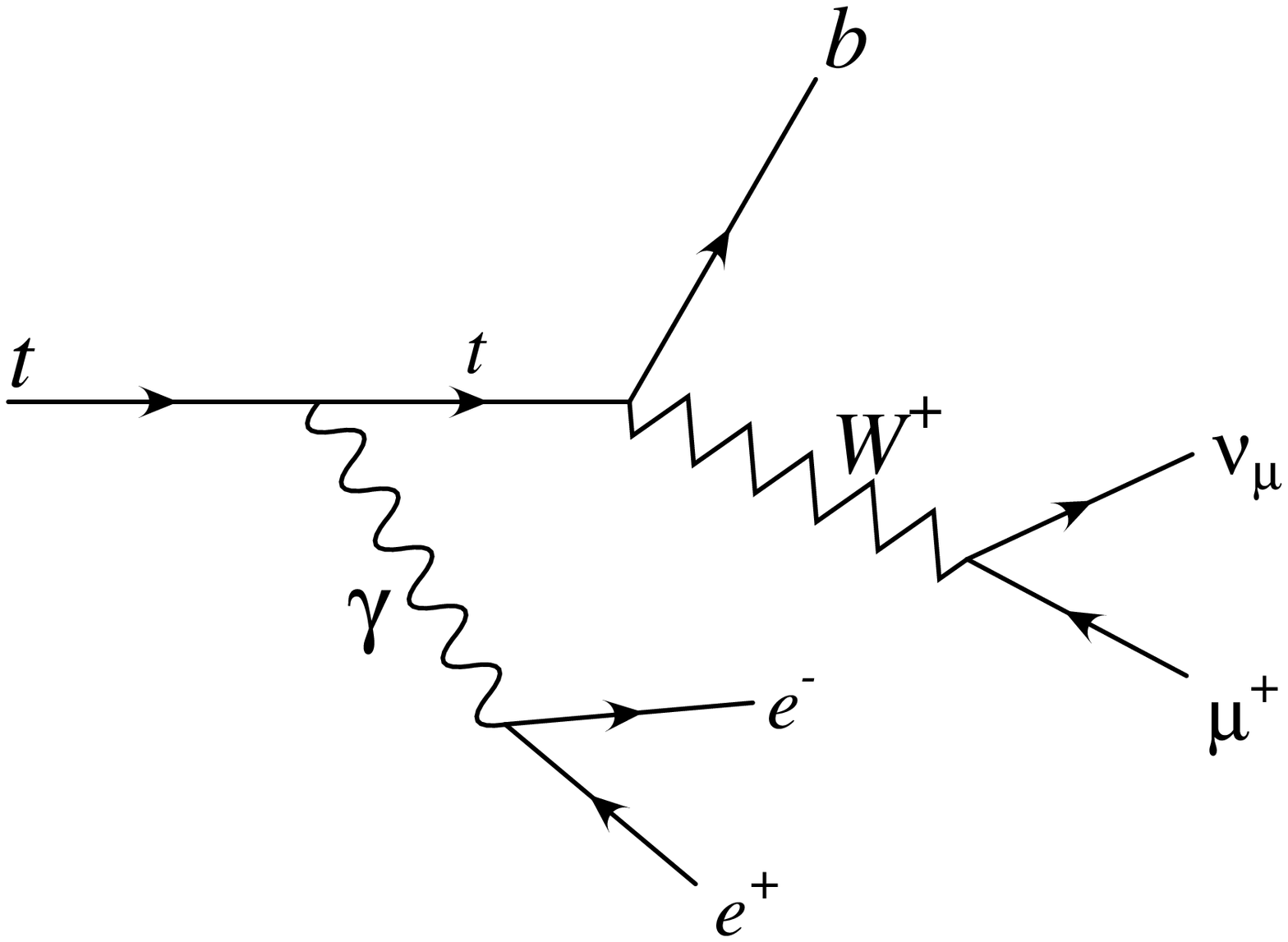}
\vspace*{-5cm}
\caption[]{Irreducible background contributions to the process
$t \rightarrow WbZ \rightarrow \mu^{+}\nu_\mu b e^{+}e^{-}$
where a photon appears instead of a $Z$ boson.
These diagrams may be suppressed by cutting on the invariant
mass of the $e^{+}e^{-}$ pair.}
\label{photon}
\end{figure}

\begin{figure}
\vskip9.5cm
\includegraphics{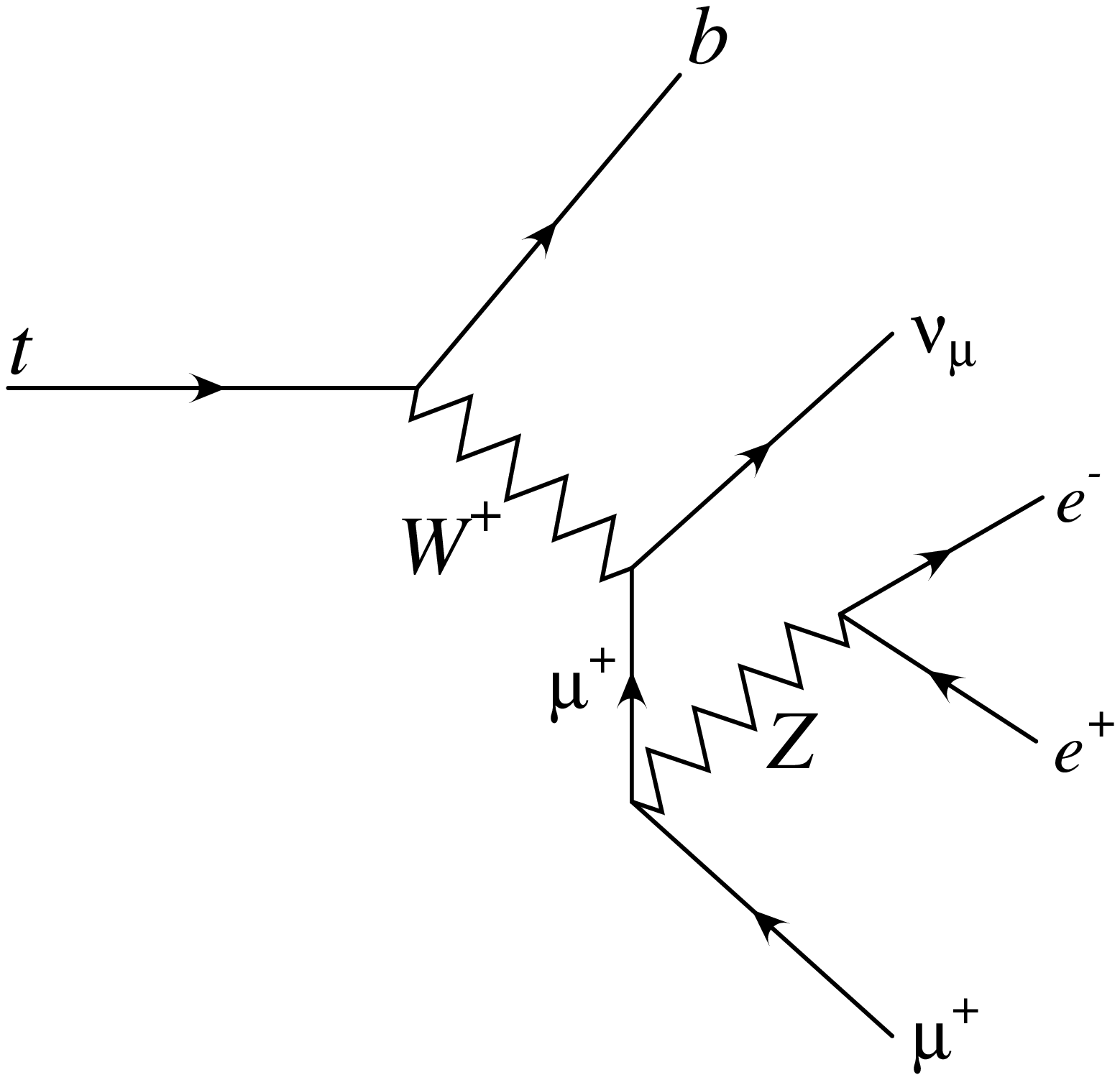}
\includegraphics{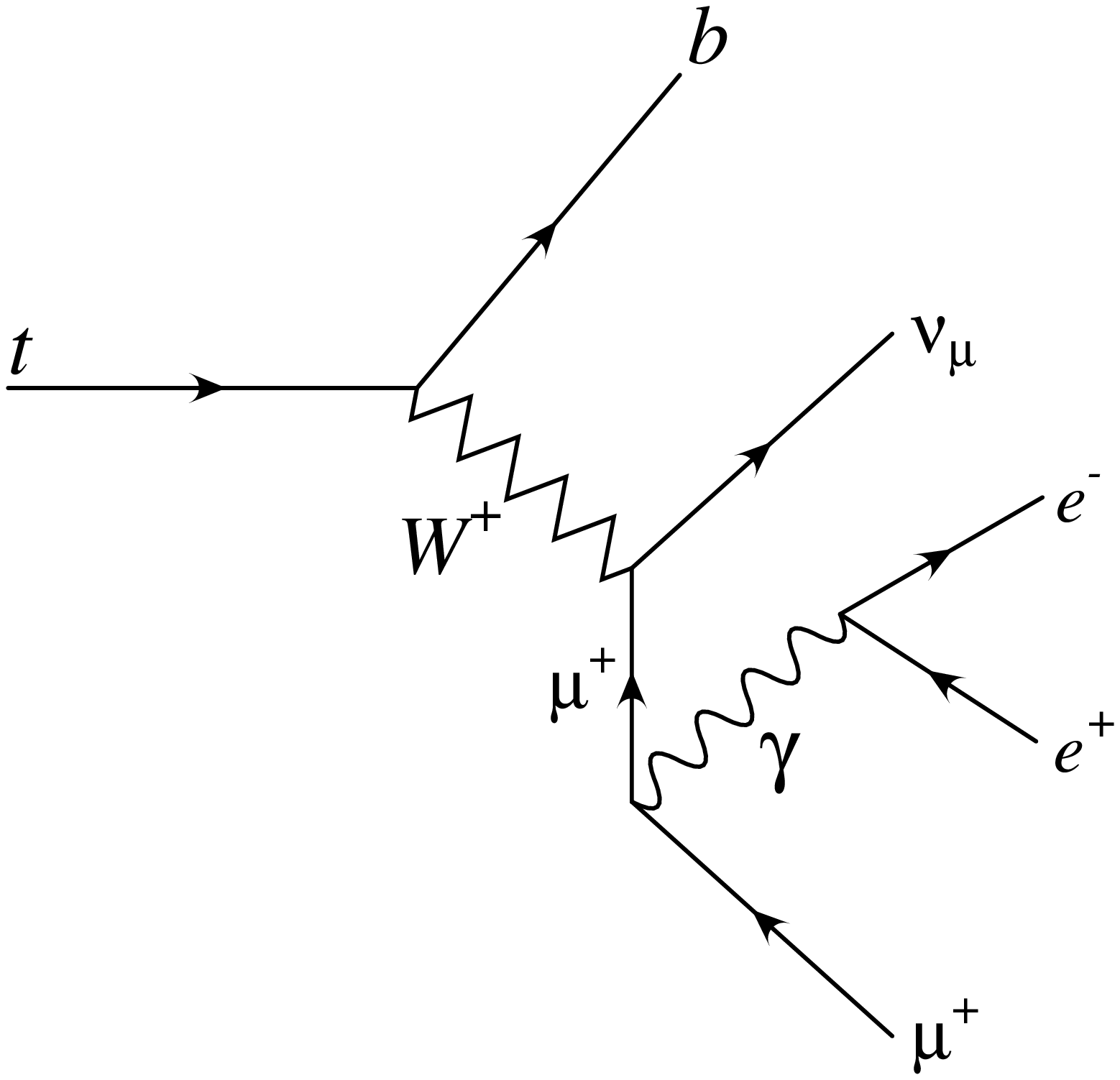}
\includegraphics{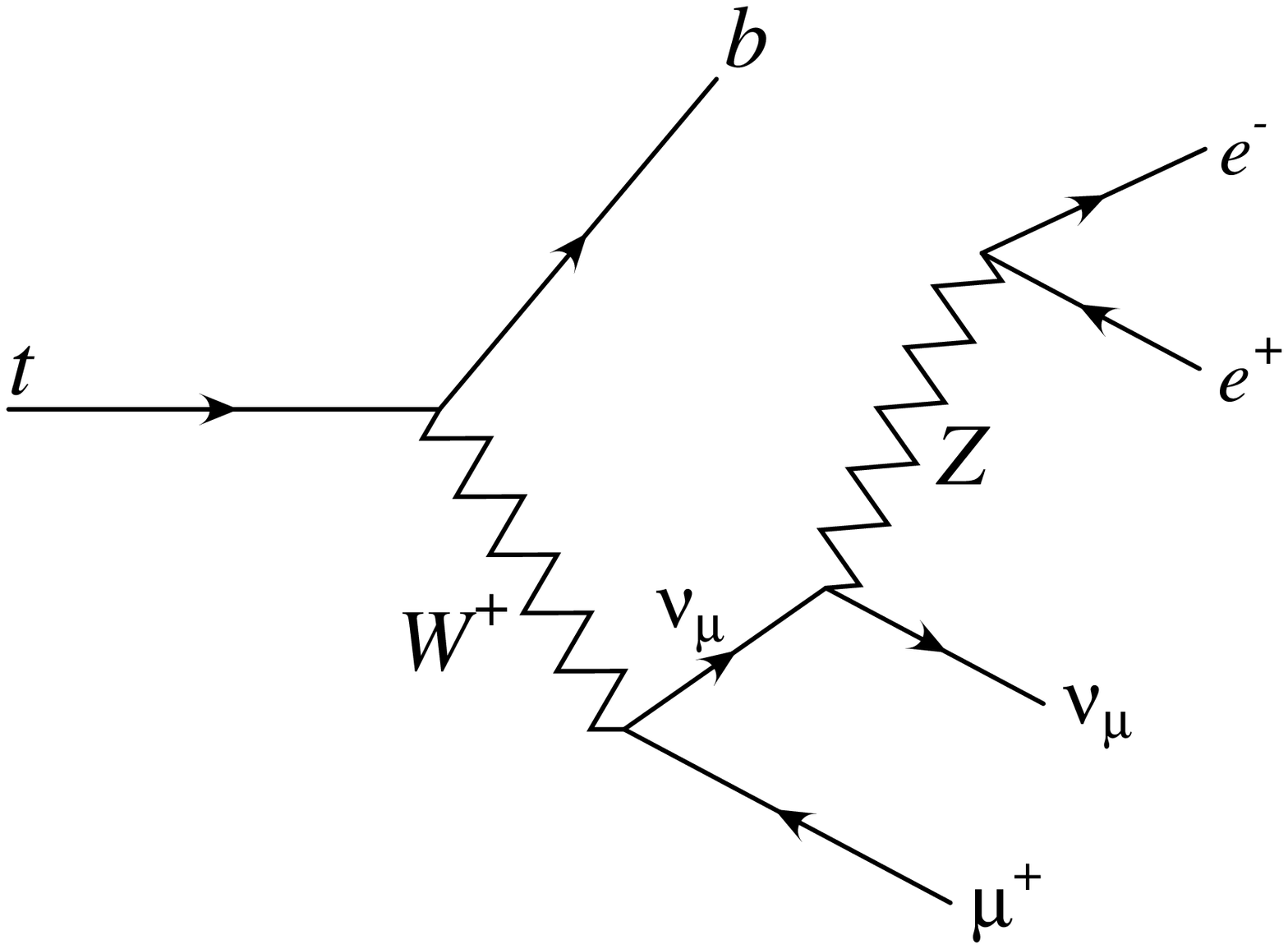}
\vspace*{-5cm}
\caption[]{Non-resonant background contributions to the process
$t \rightarrow WbZ \rightarrow \mu^{+}\nu_\mu b e^{+}e^{-}$.
These diagrams contain an extra highly off-shell propagator.}
\label{small}
\end{figure}

The first three diagrams (Fig.~\ref{resonant}) are
simply the stable particle 
diagrams for $t\rightarrow WbZ$ with the $W$ and $Z$
decays tacked on.
We refer to them as resonant diagrams since for
a heavy enough top quark, the $W$ and $Z$ are both on-shell.
These are the dominant contributions to the total width.
However, consistency demands that we include additional diagrams.
For example, everywhere a $Z$ appears in Fig.~\ref{resonant},
we must also substitute a photon (see Fig.~\ref{photon}).
These diagrams contribute to the irreducible background to the
process of interest.  Their effect is minimal once we insist that
the $e^{+}e^{-}$ pair we observe reconstructs to a $Z$ boson.
In our calculation, we have required that the $e^{+}e^{-}$ pair
mass be at least $0.8M_Z$.
Finally, we also have the diagrams in Fig.~\ref{small}.
The kinematics of these diagrams are such that they are suppressed:
compared to the primary set of Fig.~\ref{resonant}, the
diagrams of Fig.~\ref{small} contain an additional highly
off-shell propagator.  Thus, they contribute very little to
the rate.

                %%%%%%%%%%%%%%%%%%%%%%%%%%%%
                %%                        %%
                %%        FIGURE 2        %%
                %%                        %%
                %%%%%%%%%%%%%%%%%%%%%%%%%%%%

                %%%%%%%%%%%%%%%%%%%%%%%%%%%%
                %%                        %%
                %%        FIGURE 3        %%
                %%                        %%
                %%%%%%%%%%%%%%%%%%%%%%%%%%%%

                %%%%%%%%%%%%%%%%%%%%%%%%%%%%
                %%                        %%
                %%        FIGURE 4        %%
                %%                        %%
                %%%%%%%%%%%%%%%%%%%%%%%%%%%%
\begin{figure}
\vskip12.0cm
\includegraphics{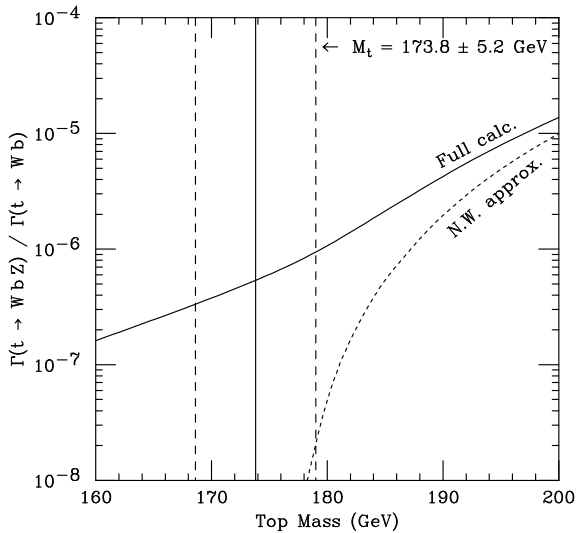}
\vspace*{-5cm}
\caption[]{The ratio
$\Gamma(t\rightarrow WbZ)/\Gamma(t\rightarrow Wb)$
as a function of the top quark mass, with an $e^{+}e^{-}$
invariant mass cut of $0.8M_Z$.  The solid curve is the
full calculation including the $W$ and $Z$ width effects,
while the dotted curve is the narrow width approximation.
For reference, the top quark mass range from the 1998
Review of Particle Properties\ts\protect\cite{PDG} is indicated.}
\label{WbZplot}
\end{figure}

Fig.~\ref{WbZplot} shows our results\ts\cite{FiniteWidth} for 
both the full calculation
as well as the so-called ``narrow-width approximation,'' which
is defined by forcing the $W$ and $Z$ to be on mass shell.
Hence, the narrow-width rate goes to zero precisely at threshold.
On the other hand, for very large top mass, both the full
and narrow-width calculations reproduce the stable particle
results presented in Ref.~\ref{WbZ1}.\footnote{Ref.~\ref{WbZ2}
presents results for $t\rightarrow WbZ$ which are in disagreement
with our results as well as those in Ref.~\ref{WbZ1}.}

Compared to the uncertainty in the top quark mass, 
the uncertainty in the $b$ quark mass is negligible.  
Corresponding to the range of masses
from the 1998 Review of Particle Properties,\cite{PDG}
we obtain
\beq
{ {\Gamma(t\rightarrow W b Z)}
\over
{\Gamma(t\rightarrow W b)} }
= (5.4^{+4.7}_{-2.0}) \times 10^{-7}.
\eeq
Thus, the Standard Model prediction for this decay is
well beyond the sensitivity of Tevatron Run II or
even Run III.  Its observation would imply new physics.

%%%%%%%%%%%%%%%%%%%%%%%%%%%%%%%%%%%%%%%%%%%%%%%%%%%%%%%%%%%%%%%%
%%
%%    t --> W b H
%%
%%%%%%%%%%%%%%%%%%%%%%%%%%%%%%%%%%%%%%%%%%%%%%%%%%%%%%%%%%%%%%%%

\begin{figure}
\vskip11.0cm
\includegraphics{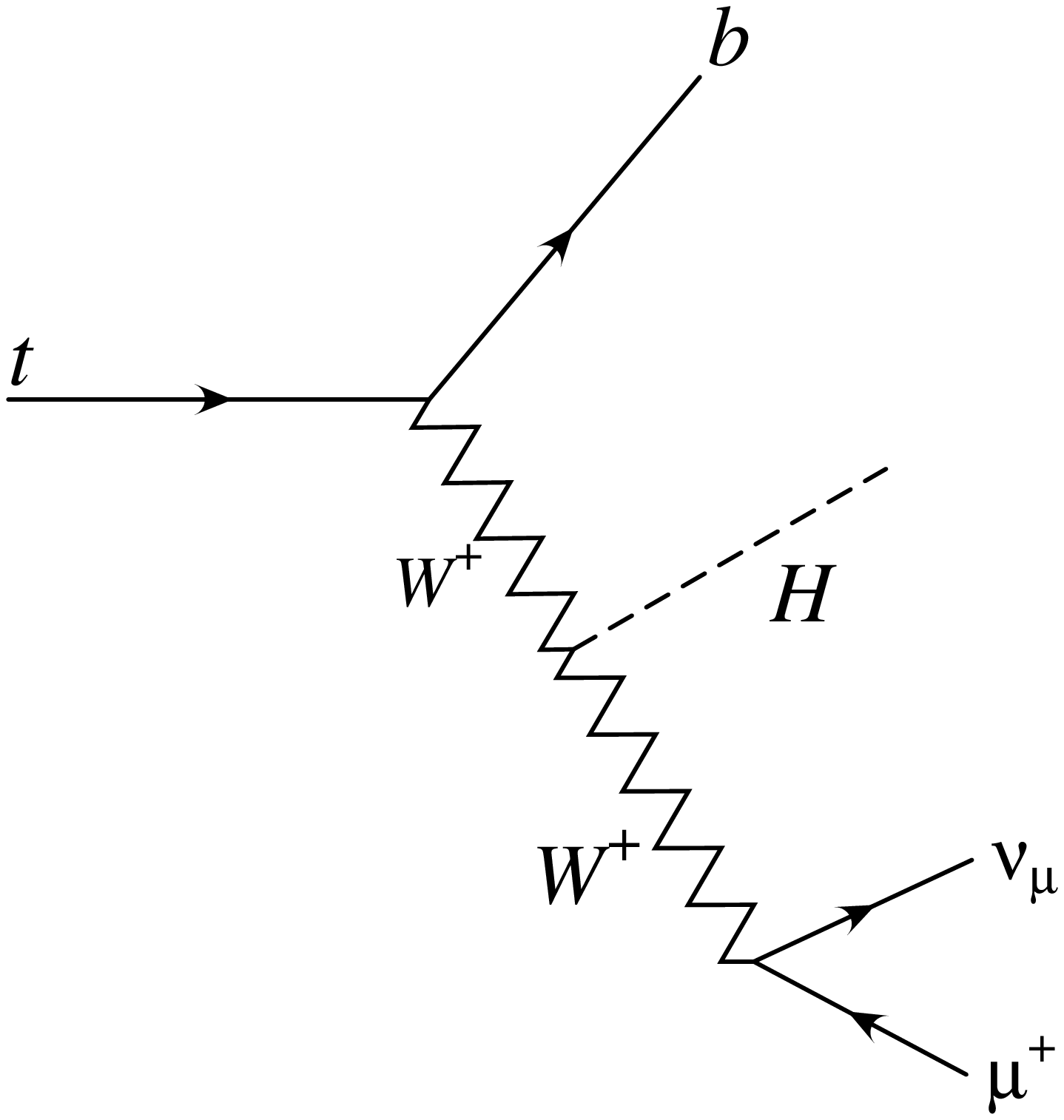}
\includegraphics{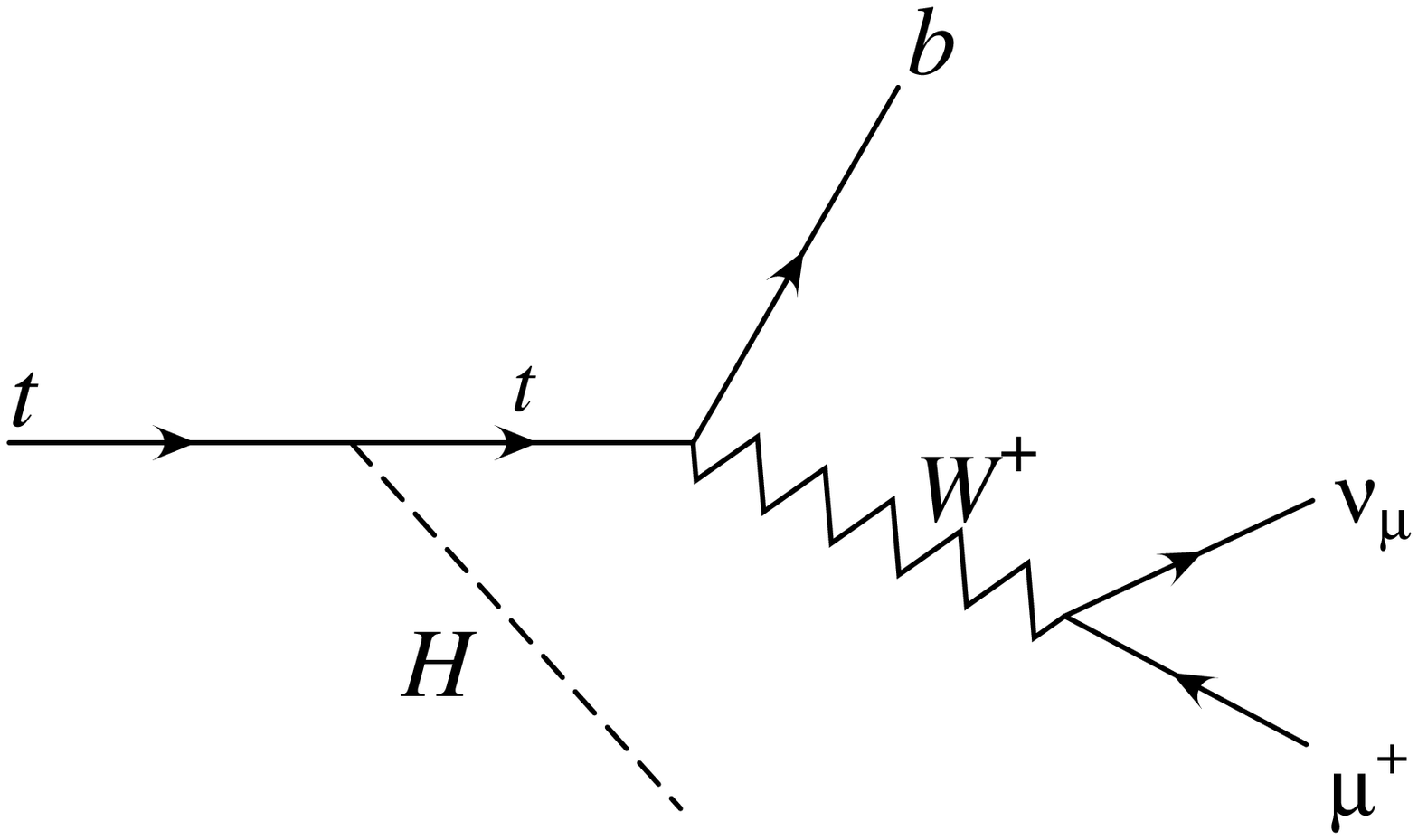}
\includegraphics{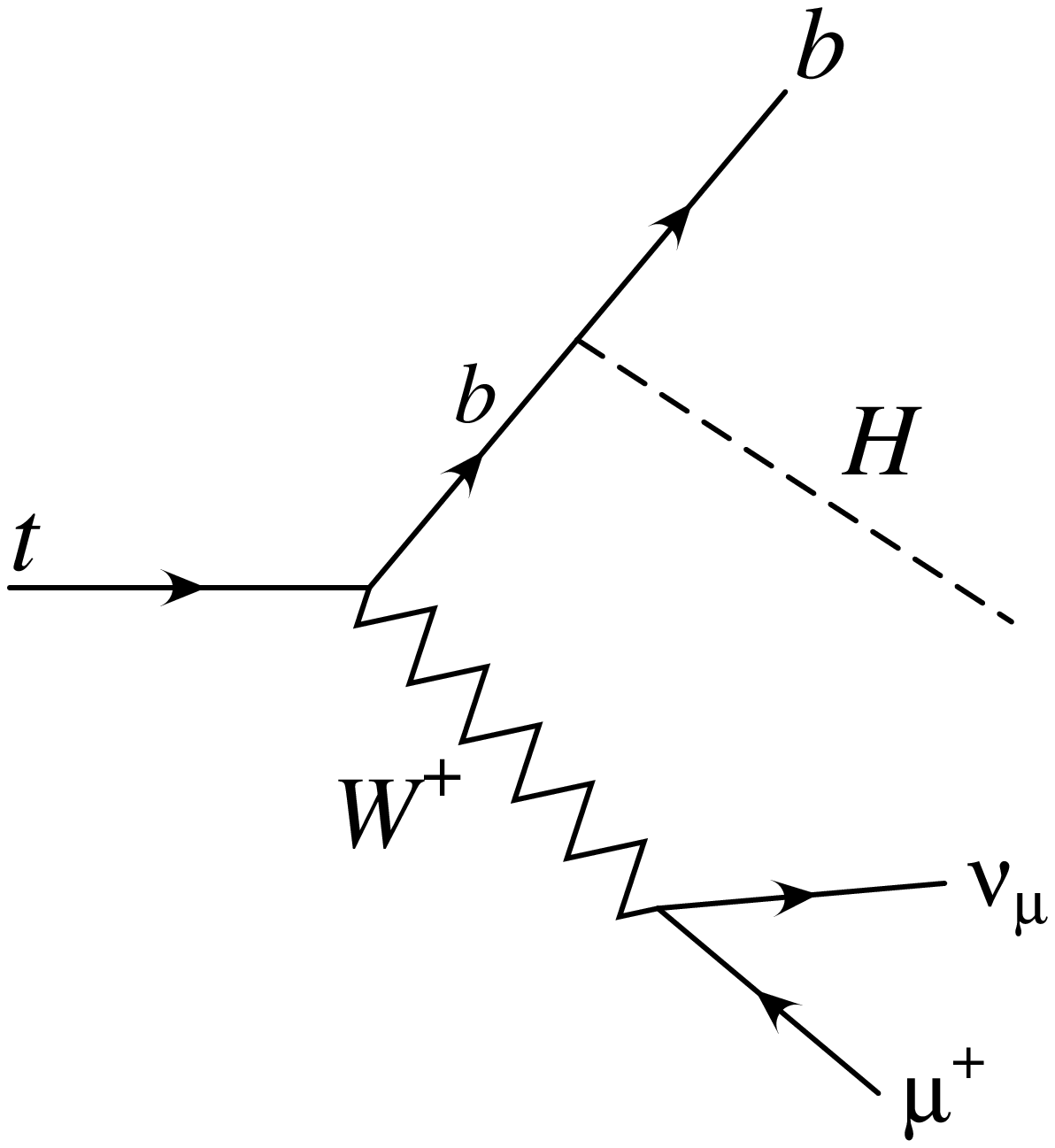}
\includegraphics{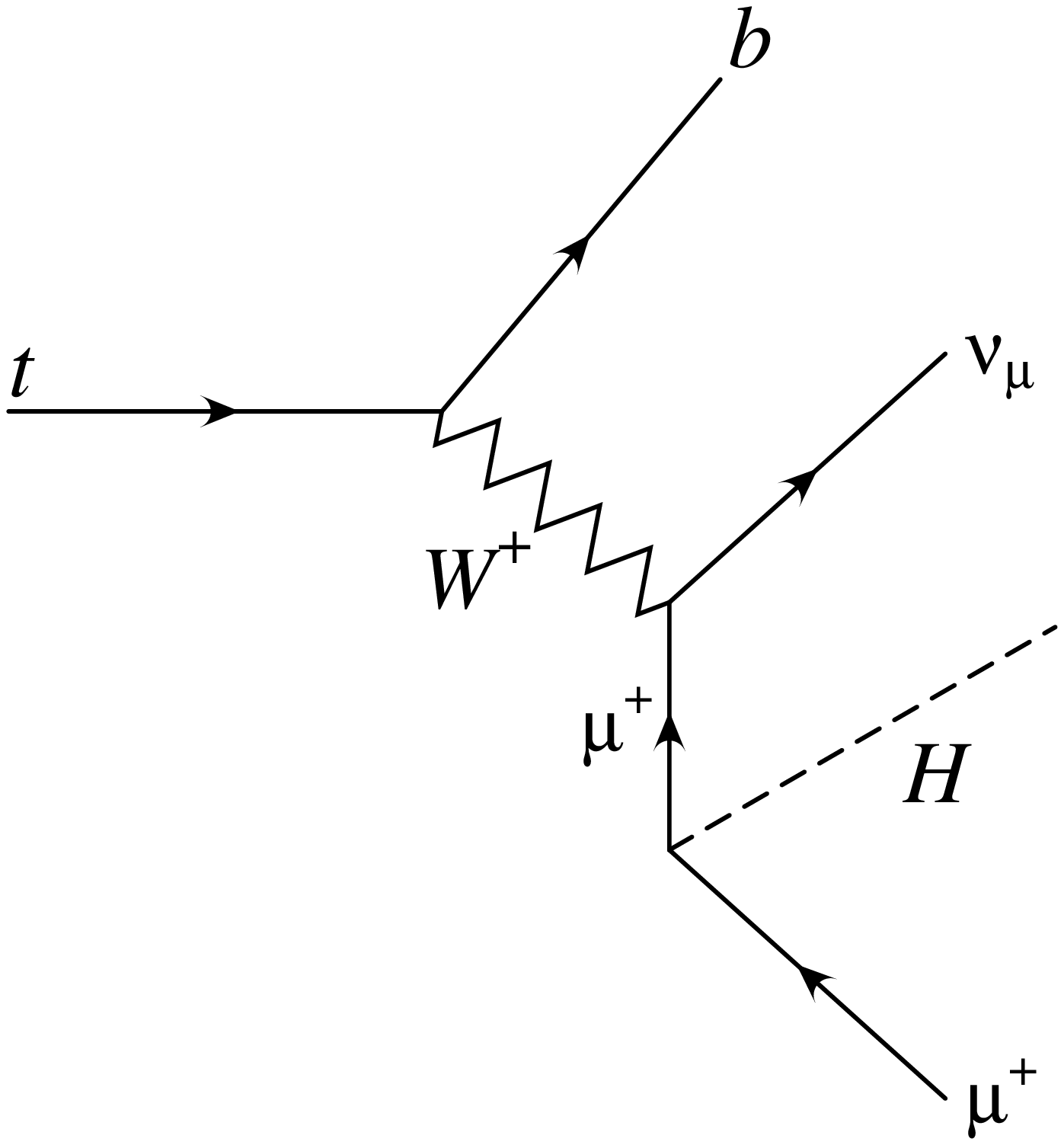}
\vspace*{-5cm}
\caption[]{Feynman diagrams for the process
$t \rightarrow WbH \rightarrow \mu^{+}\nu_\mu b H$.
Since the fourth diagram is suppressed by a tiny $\mu\mu H$
coupling {\it and}\ an off-shell $W$ propagator, we ignore it.}
\label{WbHdiagrams}
\end{figure}

\begin{figure}
\vskip12.0cm
\includegraphics{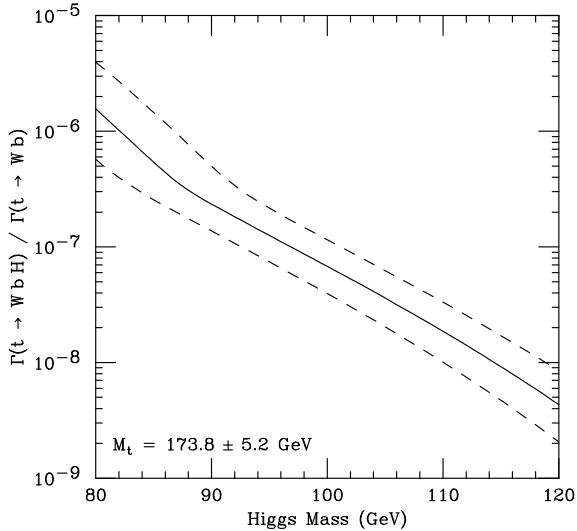}
\vspace*{-5cm}
\caption[]{The ratio
$\Gamma(t\rightarrow WbH)/\Gamma(t\rightarrow Wb)$
as a function of the Higgs boson mass.
The dotted curves indicate the values corresponding
to the top quark mass range from the 1998
Review of Particle Properties.\protect\cite{PDG}}
\label{WbHplot}
\end{figure}

We treat the decay $t\rightarrow WbH$ in analogous fashion
to $t\rightarrow WbZ$, except that now we take the Higgs
boson to be stable.\footnote{For Higgs bosons light enough
for this decay to be nearly on-shell, the corresponding
Higgs width is negligible compared to the width of the $W$
boson.}  Thus, we consider only the four diagrams shown in
Fig.~\ref{WbHdiagrams}.  
We ignore the diagram where the Higgs is emitted from
the muon since it is suppressed by a very small $\mu\mu H$ coupling
{\it and}\ an additional off-shell propagator. 
Our results appear as a function
of the Higgs mass in Fig.~\ref{WbHplot}, where the plotted error
bars only account for the (dominant) uncertainty in the top quark mass.
We have explicitly verified that our calculation agrees with
the literature\ts\cite{WbZ2,WbHlit} in the limit of large
top mass.

                %%%%%%%%%%%%%%%%%%%%%%%%%%%%
                %%                        %%
                %%        FIGURE 5        %%
                %%                        %%
                %%%%%%%%%%%%%%%%%%%%%%%%%%%%

Two of the four LEP collaborations have published 95\% C.L.
lower limits on the Higgs mass based on the 1997
run at $\sqrt{s}=183\GeV$:  
L3 finds that $M_H > 87.6\GeV$~\cite{L3Higgs}
while ALEPH reports $M_H > 87.9 \GeV$.\cite{ALEPHHiggs}
Taking into account these limits, we see that 
${\Gamma(t\rightarrow W b H)} / {\Gamma(t\rightarrow W b)} $
is at most a few times $10^{-7}$.  Once again, this rate is so
tiny that observation of this mode in Runs II or III would imply
non-Standard Model physics.

                %%%%%%%%%%%%%%%%%%%%%%%%%%%%
                %%                        %%
                %%        FIGURE 6        %%
                %%                        %%
                %%%%%%%%%%%%%%%%%%%%%%%%%%%%

                %%%%%%%%%%%%%%%%%%%%%%%%%%%%
                %%                        %%
                %%        FIGURE 7        %%
                %%                        %%
                %%%%%%%%%%%%%%%%%%%%%%%%%%%%
\begin{figure}
\vskip12.0cm
\includegraphics{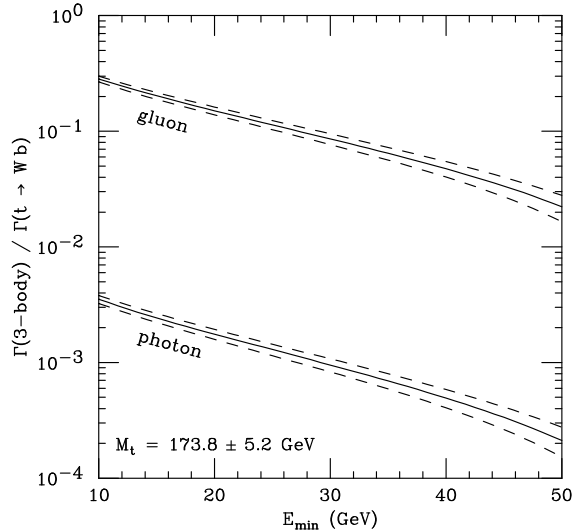}
\vspace*{-5cm}
\caption[]{The ratios
$\Gamma(t\rightarrow Wbg)/\Gamma(t\rightarrow Wb)$ and
$\Gamma(t\rightarrow Wb\gamma)/\Gamma(t\rightarrow Wb)$
as a function of the minimum gluon or photon energy
in the top quark rest frame.  The strong radiative
decay has been computed assuming fixed $\alpha_s = 0.119$.
The dotted curves indicate the values corresponding
to the top quark mass range from the 1998
Review of Particle Properties.\protect\cite{PDG}}
\label{Wbgplot}
\end{figure}

%%%%%%%%%%%%%%%%%%%%%%%%%%%%%%%%%%%%%%%%%%%%%%%%%%%%%%%%%%%%%%%%
%%
%%    t --> W b g, gamma
%%
%%%%%%%%%%%%%%%%%%%%%%%%%%%%%%%%%%%%%%%%%%%%%%%%%%%%%%%%%%%%%%%%

For completeness,
we will say a few words about the decays $t\rightarrow Wbg$
and $t\rightarrow Wb\gamma$.  These decays have been 
well-documented in Refs.~\ref{WbZ1} and~\ref{GluonPhoton}.
Both of these amplitudes are infrared divergent.  
Hence, the observed rate will depend in detail
upon issues like the 
detector resolution and (in the case of $Wbg$) the
jet isolation algorithm.
In addition,  the shift in gluon or photon energy caused
by the boost from the top quark rest frame to the lab
frame will introduce a dependence on how the tops were
originally produced.
Thus, a careful calculation of these rates would include
the full production process as well as a complete
detector simulation.
Nevertheless, we may get a feel for the behavior of these
branching ratios by considering the (idealized) situation
where we simply cut on the gluon or photon energy in
the top quark rest frame.  From Fig.~\ref{Wbgplot}
we see that these decays hold no theoretical surprises:  the
rates are approximately
\beqa
&& \Gamma(t\rightarrow Wbg) \sim 
{\cal O}(\alpha_s) * \Gamma(t\rightarrow Wb), \cr
&& \Gamma(t\rightarrow Wb\gamma) \sim 
{\cal O}(\alpha_{em}) * \Gamma(t\rightarrow Wb).
\eeqa

It is well-known that the presence of the gluonic radiative decay
(as well as initial state gluon radiation) complicates
the issue of determining the top quark mass accurately.\cite{Orr}
In fact, extra soft jets are so common an occurrence that
one could argue that there is a sense in which the decay to $Wbg$ has
been already observed, although not unambiguously isolated.
On the other hand, the decay to $Wb\gamma$ is a bit easier to
get a handle on.  
The values indicated in Fig.~\ref{Wbgplot}
suggest that evidence for this decay mode may be accessible in Run II. 

%%%%%%%%%%%%%%%%%%%%%%%%%%%%%%%%%%%%%%%%%%%%%%%%%%%%%%%%%%%%%%%%
%%
%%      REFERENCES
%%
%%%%%%%%%%%%%%%%%%%%%%%%%%%%%%%%%%%%%%%%%%%%%%%%%%%%%%%%%%%%%%%%

\end{document}